\documentclass[12pt]{article}
\usepackage{amsmath,amssymb}
\usepackage{graphicx}
\usepackage{amsmath}
\usepackage{hyperref}
\usepackage{float}
\newcommand {\be}{\begin{equation}}
 \newcommand {\ee}{\end{equation}}
 \newcommand {\bea}{\begin{array}}
 
 \newcommand {\eea}{\end{array}}

\textwidth=6.5in \hoffset=-.75in \voffset=-1in \numberwithin{equation}{section}
\numberwithin{figure}{section}

\def\0{{(0)}}
\def\1{{(1)}}
\def\2{{(2)}}

\def\<{\langle }
\def\>{\rangle }
\def\[{\left[}
\def\]{\right]}

\begin{document}
\begin{titlepage}

\vskip1cm
\begin{center}
{~\\[140pt]{ \LARGE {\textsc{Kerr/CFT from phase space formalism   }}}\\[-20pt]}
\vskip2cm

\end{center}
\begin{center}
{M. R. Setare \footnote{E-mail: rezakord@ipm.ir}\hspace{1mm} ,
M. Koohgard \footnote{E-mail: m.koohgard@modares.ac.ir}\hspace{1.5mm} \\
{\small {\em  {Department of Science,\\
 Campus of Bijar, University of Kurdistan, Bijar, Iran }}}}\\
\end{center}
\begin{abstract}
Attempts to find black hole microstates using the Hamiltonian phase space approach have been made on the Schwarzschild spacetime. Since the Schwarzschild spacetime is also in the larger family of the Kerr spacetimes, and both are asymptotically flat, the Kerr black hole is a good option for the method development. The Kerr black hole is a spinning one. We perform this analysis on the Kerr spacetime and we obtain promising results using the covariant phase space analysis. Although we have forced ourselves to use the Bondi fall-off conditions, we find the gauge degrees of freedom that could be good candidates for the black hole microstates. The charge algebra on the boundary could be a Virasoro algebra that has a different central term than the Schwarzschild black hole. The two dimensional theory on the black hole boundary is conjectured to be conformally invariant.\end{abstract}
\vspace{1cm}


\end{titlepage}

\section{Introduction}
\label{sec:intro}

In the formation of the black hole, Hawking's calculations show that the presence of a thermal spectrum seems necessary, and that this spectrum should be a unique one~\cite{a,b}. If the Hawking radiation is essentially thermal, it is inconsistent with both the unitarity principle and the basic principles of quantum mechanics and leads to "the information loss paradox"~\cite{c}. To overcome the information paradox problem, the background metric is considered as a classical field and this thermal spectrum receives corrections of some power of $\frac{1}{S}$, where $S$ is the size of the system. An important source for these corrections are the microstates which correspond to the states in the Hilbert space. These quantum corrections are considered black hole hair, and this is contrary to the semi-classical notion that black holes are hairless~\cite{d}.

A connection has been established the black hole physics and the critical phenomena in Bose-Einstein condensates \cite{d3,d4,d}. The critical point is accompanied by the Bogoliubov modes which are small excitations around the mean field value in the Bogoliubov approximation that are
degenerate with the ground state. The degeneracy of the modes is lifted by $\frac{1}{N}$-effects in the quantum calculations. These gapless modes form a subspace in the Hamiltonian phase space that is expected to have a conformal invariance. On the other hand, lifting degenerate Bogoliubov-modes by quantum corrections can also correspond to an anomaly in the conformal invariance.

The Kerr family solutions are the use of the concept of the critical point of field theory in the gravity. In pure Einstein's gravity, the Kerr family~\cite{h} are the solutions that are asymptotically flat and they could have gapless Bogoliubov-modes~\cite{i}. In~\cite{i1}, Dvali and Gomez propose a treatment to solve the black hole information paradox. In their proposal, the black hole is considered as a bound state of $N$ weakly interacting gravitons, and in quantum physics, such a bound state is the critical point of a quantum phase transition~\cite{i2}. In this picture, Hawking evaporation is described as the depletion and evaporation of the condensate of the gravitons. In the critical point of a phase transition, the interactions of the gravitions dominate their kinetic term, and at the point quantum effects are important. Here, similar to what happens in the field theory, the quantum effects cannot be ignored. It has been conjectured the Kerr black holes with extreme J spin could have two dimensional field
theories (CFTs) as their holographic dual \cite{Cast,d2,i3,i4,i5,i6,i7}.

We should mention another point here. The Hamiltonian phase space of the Kerr family includes infinite number of gapless excitations, but which of these degrees of freedom are the physical ones and can make the soft hair of the black hole? Some of these modes are physical and can shift in the phase space. The idea of finding the physical degrees of freedom, which has been raised, goes back to the question that Carlip~\cite{j} posed in his calculations, which was related to a general shape of the black hole. Carlip developed Strominger's work~\cite{k} in three dimensions of space-time and performed his calculations on the physical states of the black hole for a general state. Using the phase space formalism~\cite{l,m,n}, Averin in \cite{o} and \cite{av2} find the physical degrees of freedom in the Schwarzschild black hole. The physical microstates don't have zero energy and they should be the soft modes. There is a hidden conformal symmetry that appear in the near horizon region of the phase space \cite{d2,hac1,hac2,hac}. According to Averin in \cite{o}, the results can be considered as a first step in proving the
Schwarzchild/CFT duality, with the help of the phase space approach. We intend to extend
this approach to the Kerr black holes and let's take an important step in the Kerr/CFT duality.

So in this paper we try to address the problem of black hole hair which is defined as conserved charges in the case of Kerr spacetime. In this approach the so-called BMS-charges defined on the null boundary of the black hole spacetime, are responsible for "soft-hair" or zero-energy excitations of the background. These physical degrees of freedom can provide a statistical mechanical interpretation for black holes by representing their microstates semi-classically. Here especially we uses the covariant phase space method to analyze the phase space associated to these soft hair near a Kerr black hole solution. we first introduce the Bondi fall-off conditions~\cite{e,f,g} and by briefly introducing the Kerr spacetime, we show in which coordinates the metric can satisfy the fall-off conditions (section~\ref{sec:2}). In other words, we will explain that among the various forms of the Kerr metric, we use the appropriate form to satisfy the fall-off conditions. We then examine the phase space analysis and begin by finding diffeomorphisms and non-zero excitations of the metric (Section~\ref{sec:3}). Finding surface gauge aspects as well as examining geometry and symmetries on the black hole boundary is our next program in Section~\ref{sec:3}. The algebra of symmetry generators on the black hole boundary will be found and its central terms will be discussed (section~\ref{sec:4}). We examine whether this surface charge algebra can be used as the  $bms_{4}$-supertranslations or not. Finally in section~\ref{sec:5}, we justify that the dual theory on the boundary has the Virasoro-algebra, and this is a good reason that the microstates are the soft hair of the black hole.

\section{The Bondi fall-off conditions and the Kerr metric}
\label{sec:2}
We seek to consider the logical form of the metric at long distances in the Kerr spacetime as a asymptotically flat spacetime. According to the metric-based formalism of Bondi and Sachs~\cite{e,f} for the treatment of the Einstein's general relativity and the gravitational wave analysis, asymptotically flat spacetimes have the following metrics~\cite{p}
\begin{equation}\label{eq:1}
  ds^2=e^{2\beta}\frac{V}{r}du^2-2e^{2\beta}dudr+g_{AB}(dx^A-U^Adu)(dx^B-U^Bdu)
\end{equation}

The Bondi-Sachs coordinates $x^a=(u,r,x^A)$ are based on a familly of outgoing null hypersurfaces~\cite{q}. The above metric is obtained with the fall-off conditions which are as follows
\begin{equation}
g_{AB}dx^Adx^B=r^2\gamma_{AB}dx^Adx^B+O(r),
\end{equation}
where
\begin{equation}
\gamma_{AB}dx^Adx^B=d\theta^2+\sin^2\theta d\phi^2
\end{equation}
is the conformal $2$-metric with two degrees of freedom.

The hypersurfaces $x^0=u=const$ are null, i.e. $g^{ab}(\partial_au)(\partial_bu)=0$, so that $g^{uu}=0$. Since the two angular coordinates $x^A$, $(A, B, C, ...=2,3)$, are constant along the null rays, so that $g^{uA}=0$. Since we have the relation $g^{ab}g_{bc}=\delta^a_c$ between $g^{ab}$ and $g_{ab}$, we could find two zero components, $g_{rr}$ and $g_{rA}$, for the covariant form of the metric which is applied in~(\ref{eq:1}).
We consider these as the gauge fixing conditions.
 The $x^1=r$ coordinate is chosen to be in the gauge fixing condition, $\det(g_{AB})=r^4\sin^2\theta$, and this is also because the coordinate varies along the null rays.
 This condition on the metric determinants is also considered as another part of the gauge
fixing conditions.
 The remaining fall-off conditions are as follows
\begin{eqnarray}
  \beta &=& O(r^{-2})\nonumber \\
  \frac{V}{r} &=& -1+O(r^{-1}) \\
  U^A &=& O(r^{-2})\nonumber
\end{eqnarray}

In order to be able to extend Averin's calculations~\cite{o} to the Kerr black holes, we need to have the Kerr metric in the Bondi coordinates. The Kerr metric is the exact solution of the Einstein's equations describing a rotating, stationary and axially symmetric black hole. According to the $no- hair$ theorem~\cite{s}, all the black holes solutions of the Einstein's equations are characterized by three numbers: mass, charge, and angular momentum. The Schwarzschild black hole is the only mass dependent one, but the Kerr space-time is the spinning generalization of the Schwarzschild spherical space-time. The explicit form of the Kerr metric in the $Boyer-Lindquist$ coordinates is the following~\cite{h}:
\begin{equation}\label{KM:BL}
ds^2=-dt^2+\Sigma(\frac{dr^2}{\Delta}+d\theta^2)+(r^2+a^2)\sin^2\theta d\phi^2
+\frac{2Mr}{\Sigma}(a\sin^2\theta d\phi-dt)^2
\end{equation}
where
\begin{eqnarray}
  \Delta(r) &\equiv& r^2-2Mr+a^2\nonumber \\
  \Sigma(r, \theta) &\equiv& r^2+a^2\cos^2\theta
\end{eqnarray}

From the Kerr metric~(\ref{KM:BL}), it is obvious that the Kerr black hole depends on the two parameters: the mass parameter, $M$, and the angular momentum parameter, $a$. Since the metric does not have an explicit dependence on the time and the angle, the Kerr spacetime is \emph{stationary} and \emph{axisymmetric}, respectively. In the limit $r$ tends to infinity, this metric is reduced to the Minkowski metric, indicating that the Kerr spacetime is an \emph{asymptotically flat spacetime}. In the limit angular momentum, $a$, tends to zero, this metric is reduced to the Schwarzschild metric. The reason why the boundary of this metric is important to us is
that we want to finally show that there is a 2d CFT at the boundary.

The Boyer-Lindquist coordinates are good for many metric features, but not for being able to use it in the Bondi-Sachs gauge and the fall-off conditions. For this purpose, the Kerr metric is written in the Bondi coordinates as follows~\cite{r,t}
\begin{equation}\label{KM:met}
ds^2=(-1+\frac{2M}{r})dv^2+2dvdr+a\frac{\cos\theta}{\sin^2\theta}dvd\theta+r^2\gamma_{AB}dx^A dx^B
\end{equation}

In the above metric, compared to the metric in~\cite{r}, the following changes have been used to satisfy the fall-off conditions and the gauge conditions, assuming that the radius, $r$, is asymptotic.
\begin{eqnarray}
  g_{v\phi} &=& -\frac{2aM\sin^2\theta}{r}+O(r^{-2})\to 0\nonumber \\
  g_{v\theta} &=& \frac{a\cos\theta}{2\sin^2\theta}+\frac{a\cos\theta}{4}(8M+\frac{a}{sin^3\theta})\frac{1}{r}
  +O(r^{-2})\to \frac{a\cos\theta}{2\sin^2\theta}\nonumber \\
  g_{\theta\theta} &=& r^2+\frac{a}{\sin\theta}r+O(r^{-1}) \to r^2 \nonumber\\
  g_{\phi\phi} &=& r^2\sin^2\theta-a\sin\theta r \to r^2\sin^2\theta
\end{eqnarray}
The inverse metric has the following form
\begin{eqnarray}\label{KM: imet}
  g^{vv} &=& g^{vA}=g^{\theta\phi}=0\nonumber \\
  g^{\phi r} &=& 0\nonumber \\
  g^{vr} &=& 1-(-1+2\cos^2\theta)\frac{a^2}{r^2}\nonumber \\
  g^{r\theta} &=& -\frac{\cos\theta}{2\sin^2\theta}\frac{a}{r^2}\nonumber \\
  g^{\theta\theta} &=& \frac{1}{r^2}\nonumber \\
  g^{\phi\phi} &=& \frac{1}{r^2\sin^2\theta}\nonumber \\
  g^{rr} &=& 1-\frac{2M}{r}+\frac{1}{4r^2}(-8\cos^2\theta+4+\frac{1}{4}\frac{\cos^2\theta}{\sin^4\theta})
\end{eqnarray}
in which the $r$ component is asymptotic. To satisfy the Bondi gauge and the fall-off conditions, we consider the components approximately as follows
\begin{eqnarray}
  g^{vr} &\to& +1\nonumber \\
  g^{rr} &\to&  1-\frac{2M}{r}
\end{eqnarray}

Using the Kerr metric~(\ref{KM:met}) and its inverse~(\ref{KM: imet}), we calculate the connection coefficients that appear in the appendix~\ref{app}. We will use the approximate form of the metric~(\ref{KM:met}) in the following sections. We have ignored terms that have higher order of the inverse radius. However, the approximate form of the metric is in its general form in the Bondi gauge~(\ref{eq:1}). The effect of ignored terms on the diffeomorphisms and on the metric corrections will also be of orders of inverse radius, which can also be ignored. The approximate forms of the diffeomorphisms and the metric corrections will be computed in the following section.

\section{Surface degrees of freedom of a Kerr black hole}	
\label{sec:3}

The Bondi fall-off conditions are important in the study of the asymptotic symmetry algebra.
We can find the gauge transformations that preserve the Bondi fall-offs and the gauge fixing
conditions that mentioned in the previous section. The generators of these transformations can be considered as the generators of hidden symmetry that we are looking for.  The authors in \cite{p} have found such transformations for a general case in which there is no event horizon. They have worked in an asymptotically flat 4D spacetime and they have shown the existence of the $bms_4$-symmetry as a hidden symmetry at null infinity of the spacetime. We look for the hidden symmetry in  the Kerr spacetime in which there is the event horizon  and consider the symmetry as an enhancement in the symmetry of the spacetime.
To this end, we now consider the metric~(\ref{KM:met}) as a reference point in the solution space and we find the excitations $g_{ab}+h_{ab}$ that are primarily responsible for the black hole microstates. To achieve this, we assume that there is a Hamiltonian description of the phase space, at least at points close to the reference point. To analysis this phase space, we use the covariant phase space method~\cite{l,m}. According to the black hole uniqueness theorems, the only possible \emph{asymptotically flat} and \emph{axisymmetric} solutions of the Maxwell-Einstein equations are the diffeomorphic solutions to the Kerr family. So it is possible that the black hole microstates are in the form of the excitations $h_{ab}=\mathcal{L}_{\xi}g_{ab}$ which are the Lie derivatives with respect to the metric. These excitations involve the gauge transformations to the reference point and are diffemorphic to the Kerr space-time. Of course, not all of these excitations are  real and physical degrees of freedom, and only some of them could make a shift in the Hamiltonian phase space. This is according to the Hawking, Perry and Strominger's work~\cite{t1,t2,s1,s2} on \emph{the soft hair of the black hole}. The degrees of freedom associated with the microstates in the Schwarzschild black hole have been done by Averin~\cite{o,av2}, and we want to do that for the Kerr black hole.

To find the black hole microstates, the candidate excitations $h_{ab}$ should
preserve the Bondi gauge. The residual gauge transformations that preserve
the Bondi gauge fixing conditions in the form of the metric~(\ref{KM:met}) could be generated by vector fields satisfying the following conditions
\begin{equation}\label{Lx_rr}
  \mathcal{L}_{\xi}g_{rr} = 0
\end{equation}
\begin{equation}\label{Lx_rA}
  \mathcal{L}_{\xi}g_{rA} = 0
\end{equation}
\begin{equation}\label{Lx_AB}
  \mathcal{L}_{\xi}g_{AB}g^{BA} = 0
\end{equation}

By the usual definition of the Lie derivative, the first condition~(\ref{Lx_rr}) is expanded as follows

\begin{equation}\label{delx}
  \nabla_r\xi_r=\nabla_r\xi^v=\partial_r\xi^v=0
\end{equation}

We have used the metric compatibility condition in the first equality. The Kerr metric components~(\ref{KM:met}) and the Christoffel coefficients of the Appendix~\ref{app} are used in the $2^{nd}$ equality in (\ref{delx}). So the first component of the diffeomorphism that can be responsible
for the residual gauge
transformation can take the
following form
\begin{equation}\label{Diff-1}
  \xi^v=X(x^A,v).
\end{equation}

The $2^{nd}$ condition~(\ref{Lx_rA}) to find the residual
gauge transformation can be expanded as follows
\begin{equation}
  g_{rv}\nabla_A\xi^v+g_{AB}\nabla_r\xi^B+g_{Av}\nabla_r\xi^v=0.
\end{equation}

By the covariant derivative expansion and the Christoffel coefficients, the left hand side of the above equation has the following form
\begin{equation}\label{Diff-2}
  \partial_A\xi^v+g_{AB}\partial_r\xi^B=0.
\end{equation}

The last condition~(\ref{Lx_AB}) to find the residual gauge transformation of the Appendix \ref{app} can be  expanded as follows
\begin{equation}\label{Diff-3}
  g^{AB}\mathcal{L}_{\xi}g_{AB}=D_A\xi^A+\frac{2}{r}\xi^r=0
\end{equation}

$D_A$ denotes the covariant derivative on $S^2$ and $D^2$ is the associated Laplace-operator. By differentiation with respect to $r$-component, we find the condition~(\ref{Diff-2}) as follows
\begin{equation}\label{Diff-3-2}
  \partial_A\partial_r\xi^v+2r\gamma_{AB}\partial_r\xi^B+r^2\gamma_{AB}\partial_r^2\xi^B=0
\end{equation}

Because of the $r$-independence of $\xi^v$ in (\ref{Diff-1}), the first term of~(\ref{Diff-3-2}) will be omitted and the general solution can take the following form
\begin{equation}\label{Diff-3-gen}
  \xi^A(r,v,x^A)=X^A(v,x^A)+\frac{1}{r}Z^A(v,x^A).
\end{equation}

Substituting~(\ref{Diff-3-gen}) into~(\ref{Diff-2}), we find
\begin{equation}\label{Diff-3- Z and X}
  \partial_A\xi^v=\gamma_{AB}Z^B
\end{equation}

We substitute this result into ~(\ref{Diff-3-gen}) and we find the following form of the $3^{rd}$ component of the diffeomorphism vector
\begin{equation}\label{Diff-4}
  \xi^A=X^A+\frac{1}{r}D^AX
\end{equation}
where we have used $D_AX=\partial_AX$. By the results ~(\ref{Diff-1}),~(\ref{Diff-3}) and~(\ref{Diff-4}), we expect the components of the diffeomorphism vector field which preserve the Bondi gauge fixing conditions, can take the following form
\begin{equation}\label{diffs}
  \xi=X\partial_v-\frac{1}{2}(rD_AX^A+D^2X)\partial_r+(X^A+\frac{1}{r}D^AX)\partial_A.
\end{equation}

Here, $X=X(v,x^A)$ and $X^A=X^A(v,x^A)$ are an arbitrary scalar and arbitrary vector field on the \emph{2 sphere}, respectively. The argument we have followed to find the candidate form of the diffeomorphism
vector is based on the Kerr metric (\ref{KM:met}) and its related Christoffel coefficients in the Appendix \ref{app}.
 This form of the residual gauge transformation also justify the general form of the diffeomorphisms in \cite{p}. The authors of \cite{p} have obtained the general form of the residual gauge transformations in asymptotically flat 4-D spacetime that our result~(\ref{diffs}) is one of the special cases of that form. In addition, this
candidate form for the diffeomorphsim is in consistent with the form of the diffeomorphsim for the Schwarzschild black hole in \cite{o}. Utilizing the diffeomorphism (\ref{diffs}) and its related Lie derivative $h_{ab}=\mathcal{L}_{\xi}g_{ab}$, the non-zero excitations on the metric components could be found as follows
\begin{align}\label{h_ab}
h_{vr} &= \partial_v X-\frac{1}{2}D_BX^B-\frac{1}{2}r\partial_r(D_BX^B)\nonumber \\
h_{Av} &= -\frac{r}{2}D_AD_BX^B-\frac{1}{2}D_AD^2X+VD_AX+r^2\partial_vX_A+r\partial_vD_AX\nonumber\\
       &+\frac{a}{2}\frac{\cos\theta}{\sin^2\theta}\delta_{A\theta}(\partial_vX+\frac{M}{r^2}X
  +\frac{a}{2}\frac{\cos\theta}{\sin^2\theta}\delta_{B\theta}(D_AX^B+\frac{1}{r}D_AD^BX)\nonumber \\
h_{AB} &= r^2(D_AX_B+D_BX_A-\gamma_{AB}D_CX^C)+r(2D_AD_BX-\gamma_{AB}D^2X)\nonumber\\
       &+\frac{a}{2}\frac{\cos\theta}{\sin^2\theta}(\delta_{B\theta}D_AX+\delta_{A\theta}D_BX)\nonumber\\
h_{vv} &= \frac{M}{r}D_BX^B+\frac{M}{r^2}D^2X+2V\partial_vX-r\partial_vD_BX^B-D^2\partial_vX\nonumber\\
       &+a\frac{\cos\theta}{\sin^2\theta}\partial_vX^A\delta_{A\theta}+\frac{a}{r}\frac{\cos\theta}{\sin^2\theta}\partial_vD^AX\delta_{A\theta}
\end{align}

Although the diffeomorphsim form (\ref{diffs}) is consistent with the case of the Schwarzschild black hole \cite{o}, this is not the case with the metric excitations (\ref{h_ab}).These excitations have been modified in related to the Schwarzschild case. In general, in the phase space formalism,
the set of configurations satisfy
 the equation of motion forms the (not gauge-fixed) covariant phase space $\mathcal{F}$.
Working in the Bondi gauge corresponds to fixing a
gauge, which is an step to obtain the gauge-fixed solution space $\Gamma\subseteq\mathcal{F}$. To find the physical
excitations of the Kerr black hole, we consider just the non-zero modes of the presymplectic form related to the gauge-fixed
covariant space $\Gamma$.

By examining the Hamiltonian generators of these excitations utilizing the phase space formalism \cite{l,m,n}, we find out which of them are physical. In general, the Hamiltoniant generator $H$ of a gauge transformation $\mathcal{L}_{\xi}g_{ab}$ over a Cauchy-surface is determined by
\begin{equation}\label{dH: gen}
  \delta H[h_{ab};g_{ab}]=\int_{\Sigma}\omega[h_{ab},\mathcal{L}_{\xi}g_{ab}; g_{ab}],
\end{equation}
where $\delta H$ denotes the variation of the Hamiltonian $H$ and $\omega$ is the presymplectic current differential form. The gauge transformation generator is composed of a bulk term and a surface contribution. The former is related to the equations of motion, which is therefore omitted. The remaining on-shell term of the generator~(\ref{dH: gen}) is as follows
\begin{equation}\label{dH: Surface}
\delta H[h_{ab};g_{ab}]=-\frac{1}{16\pi}\oint_{\partial\Sigma}*F,
\end{equation}
where $F$ is the $2-form$ of the field strength over the spacetime. In this case, $\partial\Sigma$ is a cross-section of the outer event horizon of the Kerr black hole. We perform the Hamiltonian computations on the
large radius limit. The general form of the $F$ is as follows~\cite{t2}
\begin{align}\label{Fab}
  F_{ab} & = \frac{1}{2}(\nabla_a\xi_b-\nabla_b\xi_a)+(\nabla_ah^c_b-\nabla_bh^c_a)\xi_c
  +(\nabla_c\xi_ah^c_b-\nabla_c\xi_bh^c_a)\nonumber \\
 & -(\nabla_ch^c_b\xi_a-\nabla_ch^c_a\xi_b)-(\nabla_ah\xi_b-\nabla_bh\xi_a).
\end{align}

Since the $\partial\Sigma$ has fixed $v$ and $r=r_S$ and has the topology of the \emph{2-sphere} parameterized by the polar and the azimuthal coordinates, we have
\begin{equation}\label{dH:rv}
  \delta H[h_{ab};g_{ab}]=-\frac{r^2}{16\pi}\oint_{\partial\Sigma}d^2x\sqrt{\gamma}F_{rv},
\end{equation}
where $\gamma=\det\gamma_{AB}$ and the \emph{2-form} $F$ has the following form
\begin{align}\label{Frv}
F_{rv} &=\xi^A\big(\partial_rh_{Av}-\frac{2}{r}h_{Av}+\frac{\cos\theta}{\sin^2\theta}\frac{a}{r}\delta_{A\theta}h_{vr}\nonumber\\
&-\frac{a}{2r^3}\frac{\cos\theta}{\sin^2\theta}\delta_{A\theta}h^B_B\big)+\xi^v\big(-\frac{1}{r^2}D^Ah_{Av}+\frac{1}{r^2}
\partial_vh^A_A-\frac{2}{r}h_{vv}\nonumber\\
&-\frac{M}{r^4}h^A_A+\frac{2}{2r^2}\frac{\cos\theta}{\sin^2\theta}\delta_{A\theta}(\partial_rh_{Av}+
\partial_Ah_{rv})-\frac{V}{r^2}\partial_rh^A_A\big)\nonumber\\
&+\partial_r\xi^vh_{vv}+\frac{1}{r^2}D^A\xi^vh_{vA}-\frac{1}{2r^2}\partial_v\xi^vh^A_A+\xi^r
\big(\frac{2}{r}h_{vr}+\frac{1}{r^3}h^A_A\big)+\frac{1}{2r^2}\partial_r\xi^rh^A_A.
\end{align}
which is calculated using the non-zero metric exitations~(\ref{h_ab}) and~(\ref{Fab}). Except for the terms
proportional to the angular momentum
parameter $a$, the remaining terms in~(\ref{Frv}) are similar to those form in the Schwarzschild spacetime~\cite{t2,o}. Of course, these terms are important in terms of having the angular momentum $a$ because the angular momentum is an important feature of the Kerr spacetime compared to the Schwarzchild spacetime.

At this point we calculate the change of the Hamiltonian generator between $g_{ab}$ and $g_{ab}+h_{ab}$ by substituting Eq.~(\ref{Frv}) into Eq.~(\ref{dH:rv}) and we find that its form by the excitations $\xi=\xi(Y,Y^A)$ (see~(\ref{diffs})) and $h_{ab}=h_{ab}(X,X^A)$ (see~(\ref{h_ab})) is as follows:
\begin{align}\label{dH:final}
\delta H&=-\frac{r_S}{16\pi}\oint_{\partial\Sigma}d^2x\sqrt{\gamma}\bigg\{Y(1-D^2)D_BX^B+D_BY^B(D^2-1)X\bigg\}_{r=r_S}\nonumber\\
 &-\frac{r^2_S}{16\pi}\oint_{\partial\Sigma}d^2x\sqrt{\gamma}\bigg\{Y(-\frac{a}{2r^2}\frac{\cos\theta}{\sin^2\theta}\delta_{A\theta})(D_AD_BX^B)\nonumber\\
 &+Y\big[(-\frac{a}{2r^2}\partial_{\theta}(\frac{\cos\theta}{\sin^2\theta})(2D_BX^B\delta_{B\theta}-D_BX^B)-\frac{1}{2}D^2\partial_rD_BX^B\big]\nonumber\\
 &+D_AY^A\big[(\frac{a}{2}\frac{\cos\theta}{\sin^2\theta})(\frac{2}{r}X^B\delta_{B\theta}-\frac{3}{2r^2}D^BX\delta_{B\theta})-\frac{1}{2}r\partial_rD_BX^B\big]\nonumber\\
 &+Y^A\big[(\frac{a}{2}\frac{\cos\theta}{\sin^2\theta})(-\frac{1}{r}D_BX^B\delta_{A\theta}+\partial_rD_BX^B\delta_{A\theta})+\frac{a}{r}\partial_{\theta}(\frac{\cos\theta}{\sin^2\theta})X^B\delta_{B\theta}\delta_{A\theta}\big]\bigg\}_{r=r_S}
\end{align}

The first term in~(\ref{dH:final})  is common with its form the Schwarzschild spacetime case in \cite{o} but the remaining terms are specific to the Kerr black holes geometry.

The variation $\delta H$ in (\ref{dH:final}) contain the gauge excitations $g_{ab}$ which are the non-zero presymplectic form modes.
Because the $\delta H$ in (\ref{dH:final}) is a shift in the Hamiltonian generator, its constituents are the surface gauge excitations.
As in the Schwarzschild case in \cite{o}, all the v-dependence can be omitted in the Hamiltonian variation in (\ref{dH:final}) in the Kerr
black hole. Given the $\delta H$ in (\ref{dH:final}), we consider the non-zero excitations as follows
\begin{eqnarray}\label{asp:1st}
X &=& X(x^A)\nonumber\\
X^A &=& X^A(x^B).
\end{eqnarray}
where the scalar $X$ and the vector field $X^A$ are the physical excitations which are on the $S^2$. They can be
considered as the candidates for the gauge degrees of freedom of the Kerr black hole with the metric (\ref{KM:met}). We call
the excitations (\ref{asp:1st}) as the gauge aspects. It is needed to describe the geometry and the physics of the candidate
gauge aspects clearly.

Here, there is a point to note on the computation of the Hamiltonian variation in (\ref{dH:final}).To calculate the variation of the Hamiltonian generator in (\ref{dH:final}), we have used \emph{2-Form}~(\ref{Frv}) that is defined in terms of the diffeomorphisms~(\ref{diffs}) and the metric corrections~(\ref{h_ab}). Because the metric is approximate, the pre-symplectic form (\ref{Frv}) is also approximate. If we use the exact form of the metric, some higher order $r$-inverse terms may be added to the integrand of~(\ref{dH:final}). Since we have chosen the gauge aspects based on the form of the integrand terms, adding the ignored terms does not change the gauge aspects.  This is because the possible added terms are some functions of the selected gauge aspects and their derivatives. We determine the form of the gauge aspects~(\ref{asp:1st}) according to the integrand of~(\ref{dH:final}). The ignored terms have no effects on the choice of gauge aspects. The gauge aspects are important in counting the black hole microstates. Since the ignored terms have no effect on the choice of the gauge aspects, the effects of these terms on counting microstates and obtaining black hole entropy is negligible.

Utilizing the statement in \cite{p}, the diffeomorphsim $\xi$ in (\ref{diffs}) represents a faithful representation of
the $bms_4$-supertranslation. For example, we consider an especial case of the gauge degrees as follows
\begin{eqnarray}\label{bms01}
  X &=& f(x^A),\nonumber\\
  X^A &=& 0,
\end{eqnarray}
where $f$ is a function on $S^2$. This is an especial case of diffeomorphisms  in (\ref{diffs}) and can be considered as usual $bms_4$-supertranslations \cite{p}. The $bms_4$-supertranslations are the degeneracy of the vacuum but they are not responsible for the physical microstates of the black hole \cite{str01}. The authors in \cite{dval} consider some type of the asymptotic symmetry that are not part of the $bms_4$-supertranslation. They
claim the latter symmetry is an enhanced symmetry at the event horizon of the black hole and apply this claim to the
Schwarzschild black hole. The authors in \cite{dval} have argued this enhanced symmetry can be found in the general type of the
black holes. We want to extend this claim to the Kerr black hole case (\ref{KM:met}) and we find the enhancement in the asymptotic
symmetry of the black hole which are the responsible for the black hole physical microstates. To this end, we can find there is the vector field $X^A$ on $S^2$ can be decomposed as follows \cite{o}
\begin{equation}\label{dec01}
  X^A=Y^A-D^Ag,
\end{equation}
where $D_AY^A=0$. Using the Helmholtz-Hodge decomposition (\ref{dec01}), the gauge aspects (\ref{asp:1st}) can be found in the following form
\begin{eqnarray}\label{bms02}
  X &=& f \nonumber\\
  X^A &=& Y^A-D^Ag,
\end{eqnarray}
where $f$ and $g$ are two scalars on $S^2$. Substituting the following choice of $(X,X^A)$ into the diffeomorphisms (\ref{diffs})
\begin{eqnarray}\label{bms03}
  X &=& f, \nonumber \\
  X^A &=& -\frac{1}{r_S}D^Af,
\end{eqnarray}
the diffeomorphisms at the cross-section $\partial\Sigma$ which is at the Kerr black hole event horizon has the following form
\begin{equation}\label{dif00}
  \xi|_{r=r_S}=f\partial_v.
\end{equation}

This can represent an enhanced asymptotic symmetry
that can be called the event horizon supertranslation
which is similar to the $bms_4$-supertranslation in the limit
$r_S\to\infty$. It can be seen that the degrees of freedom of the Kerr black hole can be considered by the $bms_4$-supertranslations and the event horizon supertranslations. The supertranslation (\ref{bms03}) is in the form of (\ref{bms02}) and contains a pure $bms_4$-supertranslation. These are the vacuum degeneracies. The candidates for the black hole microstates can be found, by subtracting them, as follows to get ride of the gravitational vacuum degeneracy
\begin{eqnarray}\label{Am01}
  X &=& 0 \nonumber \\
  X^A &=& Y^A-D^Ag.
\end{eqnarray}

So the gauge excitations (\ref{asp:1st}) contain the excitations (\ref{Am01}) which are the candidates for the black hole microstates. The establishment of candidates of the Kerr black hole microstates is based on the symplectic reasoning and the result is similar to the Schwarzschild black hole result in \cite{o,av2}. The only difference is that it is not possible to omit the $Y^A$ term  in the enhanced modes (\ref{Am01}).

\section{Surface charge algebra}
\label{sec:4}
In the section~\ref{sec:3}, we examined the surface degrees of freedom of the Kerr black hole. In this section, we examine the surface charge algebra, and before that we need to consider how gauge-fixing takes place in the formalism of the covariant phase space. A gauge excitation $h_{ab}$ is not a gauge-fixed one and can include the symplectic zero-modes. If these zero-modes are left out of the excitation $h_{ab}$, we get a gauge-fixed excitation $\tilde{h}_{ab}$. We find a mapping $h_{ab}\to\tilde{h}_{ab}$ that can act as an projection operator between $h_{ab}$ and $\tilde{h}_{ab}$ and fix the gauge. We then construct such an operator that can map $h_{ab}$ to $\tilde{h}_{ab}$ by separating the zero modes. $\mathcal{F}$ and $\Gamma$ are the not gauge-fixed solution space and the gauge fixed solution space of the theory,
respectively, as mentioned in the previous section. The gauge excitations are placed in the tangent spaces of $\mathcal{F}$ and $\Gamma$. The mapping between $h_{ab}\to\tilde{h}_{ab}$ corresponds to a map between the related tangent spaces.

To obtain the projection operator, we consider the Hamiltonian generators based on the surface degrees of freedom. Given the gauge excitations $h_{ab}=\mathcal{L}_{\xi}g_{ab}$ in terms of the vector field components $\xi^a$, we find the integrand ingredients of the Hamiltonian variation (\ref{Fab})
in terms of the $\xi^a$ as follows
\begin{align}\label{h:on Xi}
h^A_A &=(a\frac{\cos\theta}{\sin^2\theta})D^AX\delta_{A\theta}=(a\frac{\cos\theta}{\sin^2\theta})D^A\xi^v\delta_{A\theta}\nonumber\\
h_{Av} &=-\frac{r}{2}D_AD_B+VD_A\xi^v+r^2\partial_v\xi_A-\frac{1}{2}D_B\xi^B\delta_{A\theta}\nonumber\\
&+\frac{a}{2}\frac{\cos\theta}{\sin^2\theta}\delta_{A\theta}(\partial_v\xi^v+\frac{M}{r^2}\xi^v-\frac{1}{2}D_B\xi^B)
+\frac{a}{2}\frac{\cos\theta}{\sin^2\theta}\delta_{B\theta}D_A\xi^B\nonumber\\
\partial_rh_{Av} &=-\frac{1}{2}D_AD_B\xi^B-\frac{r}{2}\partial_rD_AD_B\xi^B-\frac{2M}{r^2}D_A\xi^v
+2r\partial_v\xi_A+r^2\partial_r\partial_v\xi_A\nonumber\\
&-\frac{1}{2}\partial_rD_B\xi^B
-\frac{a}{4}\frac{\cos\theta}{\sin^2\theta}\delta_{A\theta}\partial_rD_B\xi^B+
\frac{a}{2}\frac{\cos\theta}{\sin^2\theta}\delta_{B\theta}\partial_rD_A\xi^B\nonumber\\
D^Ah_{Av} &=-\frac{r}{2}D_BD^2\xi^B+VD^2\xi^v+r^2\partial_vD^A\xi_A-\frac{1}{2}D^AD_B\xi^B\delta_{A\theta}\nonumber\\
&+\frac{a}{2}\partial_\theta(\frac{\cos\theta}{\sin^2\theta})\delta_{A\theta}(\partial_v\xi^v+\frac{M}{r^2}\xi^v-\frac{1}{2}D_B\xi^B)\nonumber\\
&+\frac{a}{2}\frac{\cos\theta}{\sin^2\theta}\delta_{A\theta}(\partial_vD^A\xi^v+\frac{M}{r^2}D^A\xi^v-\frac{1}{2}D^AD_B\xi^B)\nonumber\\
&+\frac{a}{2}\partial_\theta(\frac{\cos\theta}{\sin^2\theta})\delta_{B\theta}\delta_{A\theta}D_A\xi^B
+\frac{a}{2}\frac{\cos\theta}{\sin^2\theta}D^2\xi^B\nonumber\\
\partial_vh^A_A &=a\frac{\cos\theta}{\sin^2\theta}\partial_vD^A\xi^v\delta_{A\theta}\nonumber\\
\partial_rh^A_A &=0
\end{align}
where we have used the excitations (\ref{h_ab}), the diffeomorphism components (\ref{diffs}) and the metric (\ref{KM:met}).

For an arbitrary vector field $\xi=\xi(X,X^A)$ and the gauge excitations and their derivatives represented in~(\ref{h:on Xi}), we find the Hamiltonian generators as follows
\begin{align}\label{dH: in proj-1}
\delta H &=-\frac{r^2}{16\pi}\oint_{\partial\Sigma}d^2x\sqrt{\gamma}\bigg\{X^A\bigg[r^2\partial_r\partial_v\xi_A
-\frac{1}{2}\partial_rD_B\xi^B
-\frac{a}{4}\frac{\cos\theta}{\sin^2\theta}\delta_{A\theta}\partial_rD_B\xi^B\nonumber\\
&+D_B\xi^B\delta_{A\theta}-\frac{a}{r}\frac{\cos\theta}{\sin^2\theta}\delta_{A\theta}(\frac{M}{r^2}\xi^v+
\frac{1}{2}D_B\xi^B)\nonumber\\
&-\frac{a}{r}\frac{\cos\theta}{\sin^2\theta}\delta_{A\theta}\partial_r\xi^r+
\partial_{\theta}(\frac{\cos\theta}{\sin^2\theta})\delta_{B\theta}\delta_{A\theta}\partial_r\xi^B\nonumber\\
&+\frac{a}{r}\partial_{\theta}(\frac{\cos\theta}{\sin^2\theta})\delta_{B\theta}\delta_{A\theta}\xi^B\bigg]
+D_AX^A\bigg[-\frac{1}{2}D_B\xi^B+\frac{r}{2}\partial_rD_B\xi^B\nonumber\\
&+\frac{2M}{r^2}\xi^v-\frac{a}{2}\frac{\cos\theta}{\sin^2\theta}\delta_{B\theta}\partial_r\xi^B+\frac{2V}{r}\xi^v
+\frac{a}{r}\frac{\cos\theta}{\sin^2\theta}\delta_{B\theta}\xi^B\nonumber\\
&-\partial_v\xi^v+\partial_r\xi^r+D_B\xi^B\bigg]+X\bigg[\frac{1}{2r}D_BD^2\xi^B+\frac{4V}{r}D^2\xi^v
-\frac{2M}{r^2}D_B\xi^B\nonumber\\
&+\frac{a}{r^2}\frac{\cos\theta}{\sin^2\theta}\partial_vD^A\xi^v\delta_{A\theta}-\frac{4V}{r}\partial_v\xi^v
-\frac{2a}{r}\frac{\cos\theta}{\sin^2\theta}\partial_v\xi^A\delta_{A\theta}\nonumber\\
&+\frac{a}{2r^2}\frac{\cos\theta}{\sin^2\theta}\delta_{A\theta}(-\frac{1}{2}D_AD_B\xi^B
-\frac{r}{2}\partial_rD_AD_B\xi^B+2r\partial_v\xi_A+r^2\partial_r\partial_v\xi_A\nonumber\\
&+\partial_A\partial_v\xi^v-\partial_r\partial_r\partial_A\xi^r-\partial_A\partial_B\xi^B)
-\frac{a}{r^2}\frac{\cos\theta}{\sin^2\theta}\delta_{A\theta}(\partial_vD^A\xi^v-D^A\partial_r\xi^r)\nonumber\\
&-\frac{a}{r^2}\partial_{\theta}(\frac{\cos\theta}{\sin^2\theta})(\partial_v\xi^v-\partial_r\xi^r)
+\frac{1}{2}\partial_rD^2D_B\xi^B-rD^A\partial_r\partial_v\xi_A\nonumber\\
&-\frac{1}{r}D^AD_B\xi^B\delta_{B\theta}+\frac{1}{r}D^2D_B\xi^B\bigg]
+\partial_vX\bigg[\frac{-a}{2r^2}\frac{\cos\theta}{\sin^2\theta}D^A\xi^v\delta_{A\theta}\bigg]\bigg\}
\end{align}
By decomposition of the vector field $\xi^A$ on $S^2$ as follows
\begin{equation}\
  \xi^A=\tilde{\xi}^A+D^Ah,
\end{equation}
where $h$ is a scalar on $S^2$ and $\tilde{\xi}^A$ is divergence-free, we have~(\ref{dH: in proj-1}) as follows
\begin{align}\label{dH: in proj-2}
\delta H &= -\frac{r^2}{16\pi}\oint_{\partial\Sigma}d^2x\sqrt{\gamma}\bigg(X^A\big(D_BY^B\delta_{A\theta}+\frac{1}{r}D^2Y\delta_{A\theta}+\frac{a}{r}\partial_{\theta}(\frac{\cos\theta}{\sin^2\theta})Y^B\delta_{B\theta}\big)\nonumber\\
&+D_AX^A\big(\frac{r}{2}\partial_rD_BY^B+\frac{2V}{r}Y+\frac{a}{r}\frac{\cos\theta}{\sin^2\theta}Y^B\delta_{B\theta}+\frac{2M}{r^2}Y\big)\nonumber\\
&+X\big(\frac{1}{r}D^2D_BY^B-\frac{1}{r}D^AD_BY^B\delta_{A\theta}-\frac{2M}{r^2}D_BY^B\big)\bigg)
\end{align}

In obtaining~(\ref{dH: in proj-2}) from ~(\ref{dH: in proj-1}), the gauge transformation $\xi=\xi(Y,Y^A)$ considered to be an excitation of the surface degrees of freedom with gauge aspects $Y$ and $Y^A$ where these are just on $S^2$ and they are independent of $v$ and $r$. In addition, we have the following relations for the $\xi$ components that are used to obtain~(\ref{dH: in proj-2})
\begin{align}
\partial_v\xi^v &=\partial_vY=\partial_vh =\partial_vD_B\xi^B=0\nonumber\\
\partial_vD^A\xi^v &=D^A\partial_r\partial_v\xi_A=\partial_vD^2\xi^v=0
\end{align}

In calculation~(\ref{dH: in proj-2}), the terms $ O(r^{-2})$ have been neglected. By comparing the coefficients of $X^A$, $D_AX^A$ and $X$ in the integrands of  the
two equations~(\ref{dH: in proj-1}) and~(\ref{dH: in proj-2}), the projection operators are obtained as follows
\begin{align}\label{pr:op}
D_BY^B+\frac{1}{r}D^2Y+\frac{a}{r}\partial_{\theta}(\frac{\cos\theta}{\sin^2\theta}) &=
+D_B\xi^B-\frac{a}{2r}\frac{\cos\theta}{\sin^2\theta}D_B\xi^B-\frac{a}{r}\partial_r\xi^r\nonumber\\
&+\frac{a}{r}\partial_{\theta}(\frac{\cos\theta}{\sin^2\theta})\xi^B\delta_{B\theta}\bigg|_{\partial\Sigma}\\
\frac{r}{2}\partial_rD_BY^B+\frac{2V}{r}Y+\frac{a}{r}\frac{\cos\theta}{\sin^2\theta}Y^B\delta_{B\theta} &= -r^2\partial_r\partial_vh\nonumber\\
&-\frac{1}{2}D_B\xi^B+\frac{r}{2}\partial_rD_B\xi^B-\frac{a}{2}\frac{\cos\theta}{\sin^2\theta}\delta_{B\theta}\partial_r\xi^B+\frac{2V}{r}\xi^v\nonumber\\
&+\frac{a}{r}\frac{\cos\theta}{\sin^2\theta}\delta_{B\theta}\xi^B-\partial_v\xi^v+\partial_r\xi^r+D_B\xi^B\bigg|_{\partial\Sigma} \label{pr:op2}\\
\frac{1}{r}D^2D_BY^B-\frac{1}{r}D^AD_BY^B\delta_{A\theta}-\frac{2M}{r^2} &= \frac{1}{2r}D_BD^2\xi^B-\frac{2M}{r^2}D_B\xi^B\nonumber\\
&-\frac{4V}{r}\partial_v\xi^v-\frac{2a}{r}\frac{\cos\theta}{\sin^2\theta}\partial_v\xi^A\delta_{A\theta}+\frac{1}{2}\partial_rD^2D_B\xi^B-rD^A\partial_r\partial_v\xi_A\nonumber\\
&-\frac{1}{r}\partial_vD^2\xi^v+\frac{1}{r}D^2\partial_r\xi^r+\frac{1}{r}D^2D_B\xi^B\bigg|_{\partial\Sigma}\label{pr:op3}
\end{align}

In the right hand side of the above projection operators, we have arbitrary gauge excitations $\xi$ do not satisfy the Bondi
gauge but in the right hand side of the above equations, we have the gauge aspects $(Y,Y^A)$ that are satisfying the
Bondi gauge. So we can claim that the above operators are the mapping between the excitations of the tangent space of
$T_{g_{ab}}\mathcal{F}$ and the excitations in the tangent space $T_{g_{ab}}\Gamma$.

Now that we have found the projection operator, we need to calculate $[\xi_1,\xi_2]$ as well to find the gauge aspects of the Lie bracket. For this purpose, we consider $\xi_1$ and $\xi_2$ as follows
\begin{eqnarray}\label{xi_1,2}
  \xi_1(X_1,X^A_1) &=& X_1\partial_{v}-\frac{1}{2}(rD_AX^A_1+D^2X_1)\partial_{r}+(X^A_1+\frac{1}{r}D^AX_1)\partial_{A}\nonumber \\
  \xi_2(X_1,X^A_2) &=& X_1\partial_{v}-\frac{1}{2}(rD_AX^A_2+D^2X_2)\partial_{r}+(X^A_2+\frac{1}{r}D^AX_2)\partial_{A}
\end{eqnarray}
where $X_i$ and $X^A_i$ are the gauge aspects that are considered as follows
\begin{eqnarray}\label{gg ascpect fr Xi_1,2}
  X_1 &=& f_1\nonumber \\
  X^A_1 &=& -D^Ag_1\nonumber \\
  X_2 &=& f_2\nonumber \\
  X^A_2 &=& -D^Ag_2
\end{eqnarray}

 We are looking for gauge aspects of the Lie-bracket $[\xi_1,\xi_2]$. Therefore, substituting the gauge aspects(\ref{gg ascpect fr Xi_1,2}) into the excitations (\ref{xi_1,2}),  we calculate its components as follows
\begin{equation}\label{br:v}
  [\xi_1,\xi_2]^v=X^A_1D_AX_2-(1\leftrightarrow 2)
\end{equation}
and
\begin{align}\label{br:r}
  [\xi_1,\xi_2]^r &=\xi_1^a\partial_a\xi_2^r-(1\leftrightarrow 2)\nonumber \\
   &=r(-\frac{1}{2}X^A_1D_AD_BX^B_2)+(\frac{1}{4}D^2X_1D_BX^B_2-\frac{1}{2}X^A_1D_AD^2X_2-\frac{1}{2}D^AX_1D_AD_BX^B_2)\nonumber \\
   & +\frac{1}{r}(-\frac{1}{2}D^AX_1D_AD^2X_2)-\frac{1}{4}\frac{\cos\theta}{\sin^2\theta}\frac{a}{r}\delta_{A\theta}
   (rD_BX^B_1+D^2X_1)(X^A_2-\frac{1}{r}D^AX_2)\nonumber \\
  &-(1\leftrightarrow 2)
\end{align}
The last component of this Lie-bracket $[\xi_1,\xi_2]$ is as follows
\begin{align}\label{br:A}
[\xi_1,\xi_2]^A & = X_1^BD_BX^A_2+\frac{1}{r^2}(\frac{1}{2}D^2X_1D^AX_2+D^BX_1D_BD^AX_2)\nonumber\\
  &+\frac{1}{r}(\frac{1}{2}D_BX^B_1D^AX_2+X_1^BD_BD^AX_2+D^BX_1D_BX^A_2)\nonumber\\
  &-(1\leftrightarrow 2)
\end{align}

The only difference in the components of the bracket $[\xi_1,\xi_2]$,~(\ref{br:v})-(\ref{br:A}), compared to the similar case in the Schwarzschild black hole calculations in~\cite{o} is related to the $r$-component of the Lie bracket in (\ref{br:r}) compared to the other components in (\ref{br:v}) and (\ref{br:A}). This component has four more terms in related to the Schwarzschild case in \cite{o,av2} that are proportional to the angular momentum. Since the Lie-bracket $[\xi_1,\xi_2]$ itself is a vector field, we replace its components ~(\ref{br:v})-(\ref{br:A}) in right hand side of the projection operators~(\ref{pr:op}), (\ref{pr:op2}) and (\ref{pr:op3}) and the projection operator for the Lie bracket can be found as follows
\begin{align}\label{1st pr}
  D_BY^B+\frac{1}{r}D^2Y+\frac{a}{r}\partial_{\theta}(\frac{\cos\theta}{\sin^2\theta})Y^B\delta_{B\theta} & =X^C_1D_BD_CX^B_2\nonumber \\
  & -\frac{a}{r}\frac{\cos\theta}{\sin^2\theta}\bigg[-\frac{r}{2}X^A_1\partial_rD_AD_BX^B_2\nonumber\\
  &+\frac{1}{4}\frac{\cos\theta}{\sin^2\theta}
  \frac{a}{r^2}\delta_{A\theta}X^A_2(D^2X_1-r^2\partial_rD_BX_1^B)\bigg]\nonumber \\
  & +\frac{a}{r}\partial_{\theta}(\frac{\cos\theta}{\sin^2\theta})\bigg[X_1^CD_CX_2^B+D^CX_1D_CX^B_2\nonumber\\
  &+\frac{1}{r}
  (\frac{1}{2}D_CX_1^CD^BX_2+X_1^CD_CD^BX_2)\bigg]\nonumber \\
  & +(\frac{1}{r}-\frac{a}{2r}\frac{\cos\theta}{\sin^2\theta})\bigg[\frac{1}{r}(\frac{1}{2}D_CX_1^CD^2X_2 \nonumber\\
  &+D^CX_1D_BD_CX_2^B)\bigg]
\end{align}

This is one of the three projection operators that we neglect $O(r^{-3})$ and $O(r^{-4})$ terms in its calculation. We have two equations for other projection operators which are given below. Neglecting the mentioned terms has been done in the next two operators as well.
\begin{align}\label{2nd pr}
  \frac{r}{2}\partial_rD_BY^B+\frac{2V}{r}Y
  &+\frac{a}{r}\frac{\cos\theta}{\sin^2\theta}Y^B\delta_{B\theta}\nonumber\\
  &=
  \frac{1}{2}X_1^CD_CD_BX_2^B+\frac{3}{2r}\bigg(\frac{1}{2}D_CX_1^CD^2X_2\nonumber \\
  & +D^CX_1D_BD_CX_2^B\bigg)-\frac{r}{4}X_1^C\partial_rD_CD_BX_2^B\nonumber\\
  & +\frac{1}{2r}\bigg(\frac{1}{2}\partial_rD_CX_1^CD^2X_2+\frac{1}{2}D_CX_1^C\partial_rD^2X_2\nonumber\\
  & +D^CX_1\partial_rD_BD_CX_2^B\bigg)-\frac{1}{2}X_1^CD_CD_BX_2^B\nonumber\\
  & -\frac{a}{2}\frac{\cos\theta}{\sin^2\theta}\delta_{B\theta}\bigg[X_1^C\partial_rD_CX_2^B-\frac{1}{r^2}(\frac{1}{2}D_CX_1^CD^BX_2\nonumber\\
  & +X_1^CD_CD^BX_2+D^CX_1D_CX_2^B)\bigg]\nonumber\\
  & +\frac{2V}{r}\bigg(X_1^BD_BX_2\bigg)+\frac{a}{r}\frac{\cos\theta}{\sin^2\theta}\delta_{B\theta}\bigg[X_1^CD_CX_2^B\nonumber\\
  & +\frac{1}{r}(\frac{1}{2}D_CX_1^CD^BX_2+X_1^CD_CD^BX_2+D^CX_1D_CX_2^B)\bigg]\nonumber\\
  & -\frac{r}{2}X_1^A\partial_rD_AD_BX_2^B+\frac{1}{4}\partial_rD^2X_1D_BX_2^B\nonumber\\
  & +\frac{1}{4}D^2X_1\partial_rD_BX_2-\frac{1}{2}X_1^A\partial_rD_AD^2X_2^B\nonumber\\
  & -\frac{1}{2}D^AX_1\partial_rD_AD_BX_2^B+\frac{1}{2r^2}D^AX_1D_AD^2X_2\nonumber\\
  & +\frac{1}{4}\frac{\cos\theta}{\sin^2\theta}\frac{a}{r^2}\delta_{A\theta}\bigg(D^2X_1D_2^A-r^2\partial_rD_BX_1^BX_2^A\bigg)
\end{align}

As expected, the second projection operator~(\ref{2nd pr}) has more terms than the first operator~(\ref{1st pr}). The last equation related to the projection operators is as follows
\begin{align}\label{3rd pr}
  \frac{1}{r}D^2D_BY^B-\frac{1}{r}D^AD_BY^B\delta_{A\theta}
  &-\frac{2M}{r}D_BY^B \nonumber\\
  & =\frac{3}{2r}D^2D_B\big(X_1^CD_CX_2^B+D^CX_1D_CX_2^B\nonumber \\
  & +\frac{1}{2}D_CX_1^CD^BX_2+X_1^CD_CD^BX_2\big)\nonumber\\
  & -\frac{2M}{r^2}D_B\big(X_1^CD_CX_2^B\big)\nonumber\\
  & -\frac{3}{4r^2}D^2D_B\big(X_1^CD_CX_2^C+\frac{1}{2}D_CX_1^CD^BX_2\nonumber\\
  & +X_1^CD_CD^BX_2+D^CX_1D_CX_2^B\big)\nonumber\\
  & +\frac{1}{r}D^2\bigg[-\frac{1}{2}X^A_1D_AD_BX_2^B-\frac{1}{2}X_1^A\partial_rD_AD_BX_2^B\nonumber\\
  & +\frac{1}{4}\partial_rD^2X_1D_BX_2^B+\frac{1}{4}D^2X_1\partial_rD_BX_2^B\nonumber\\
  &-\frac{1}{2}X_1^A\partial_rD_AD^2X_2-\frac{1}{2}D^AX_1\partial_rD_AD_BX_2^2\bigg]\nonumber\\
  & -\frac{a}{4r}\frac{\cos\theta}{\sin^2\theta}D^2(\partial_rD_BX_1^BX_2^A)\delta_{A\theta}-\big(1\leftrightarrow 2\big)
\end{align}

We find many terms in the projection operators~(\ref{1st pr})-~(\ref{3rd pr}) are not omitted, while many of these terms are zero in the Schwarzschild spacetime in \cite{o,av2}.

Generally, the Hamiltonian generators $H_X$ and $H_Y$ which are related to the symplectic symmetries $X$ and $Y$
could have the following Lie algebra \cite{o,av2}
\begin{equation}\label{genAlg}
\{H_X,H_Y\}=H_{[X,Y]}+K_{X,Y},
\end{equation}
where $K_{X,Y}$ is a central extension which is a c-number.

Substituting $X=\delta_{\xi_1}$ and $Y=\delta_{\xi_2}$, the Hamilton generators of the gauge transformations could have the following algebra, which are accompanied by a central extension, which is a $c$-number~\cite{o}
\begin{equation}\label{algebra}
  \{H_{\xi_1},H_{\xi_2}\}=H_{[\xi_1,\xi_2]}+K_{\xi_1,\xi_2}
\end{equation}
where $[\xi_1,\xi_2]$ is the Lie-bracket of vector fields on the spacetime manifold.

Considering the gauge aspects defined for the algebra~(\ref{algebra}), we get its form as follows
\begin{equation}\label{algebra:2}
  \{H_{X_1,X_1^A},H_{X_2,X_2^A}\}=H_{(Y,Y^A)}+K_{(X_1,X_1^A),(X_2,X_2^A)}.
\end{equation}

Given that the left hand side of the above equation is the variation in the Hamiltonian generators as $\big\{H_{(X_1,X_1^A)},H_{(X_2,X_2^A)}\big\}$, we obtain the central term using~(\ref{dH:final}) as follows:
\begin{align}\label{central tm}
  K_{(X_1,X_1^A),(X_2,X_2^A)} & =-\frac{r^2}{16\pi}\oint_{\partial\Sigma}d^2x\sqrt{\gamma}\Big[X_2^A(\frac{1}{r} D_BX_1^B+\frac{1}{r^2}D^2X_1-\partial_rD_BX_1^B\nonumber\\
  & +\frac{a}{4}\frac{\cos\theta}{\sin^2\theta}
  \partial_rD_BX_1^B\delta_{B\theta})\delta_{A\theta}+D_AX_2^A(\frac{r}{2}\partial_rD_BX_1^B-\frac{1}{r}D^2X_1\nonumber\\
  & +\frac{a}{r}\frac{\cos\theta}{\sin^2\theta}X_1^B\delta_{B\theta}+\frac{2M}{r}X_1-\frac{1}{2}\partial_rD^2X_1\nonumber\\
  & +\frac{5a}{4r^2}\frac{\cos\theta}{\sin^2\theta}\delta_{B\theta}D^BX_1) +X_2(\frac{1}{r}D^2D_BX_1^B-\frac{1}{r^2}D^AD_BX^B_1\delta_{A\theta}\nonumber\\
  &-\frac{2M}{r^2}D_BX_1^B+\frac{a}{2r^2}\partial_{\theta}(\frac{\cos\theta}{\sin^2\theta})
  D_BX_1^B+\frac{a}{4r^2}\frac{\cos\theta}{\sin^2\theta}D^AD_BX_1^B\delta_{A\theta}\nonumber\\
  &+\frac{1}{2}D^2\partial_rD_BX_1^B)\Big]-H_{(Y,Y^A)}[g_{ab}]
\end{align}
where we have done the following replacements
\begin{eqnarray}
  Y &\to & X_1,~~~~~X^B\to X_2^B\nonumber \\
  Y^A &\to & X_1^A, ~~~~~X\to X_2
\end{eqnarray}

We have the freedom to put all surface charges at the reference point $g_{ab}$ equal to zero, as follows
\begin{equation}\label{H=0}
  H_{(Y,Y^A)}[g^{ab}]=0
\end{equation}

Utilizing the following selection for the gauge aspects into the surface algebra,
\begin{eqnarray}\label{sel gauge aspects}
  X_1 &=& f_1\nonumber \\
  X_1^A &=& -D^Ag_1\nonumber\\
  X_2 &=& f_2\nonumber\\
  X_2^A &=& -D^Ag_2,
\end{eqnarray}
the surface charge algebra for the surface degrees of freedom of a Kerr black hole can be found as follows
\begin{equation}\label{algebra in surface}
  \{H_{f_1,g_1},H_{f_2,g_2}\}=H_{\hat{f},\hat{g}}+K(f_1,g_1;f_2,g_2)
\end{equation}
where the central term $K(f_1,g_1;f_2,g_2)$ of this algebra can be found as follows
\begin{align}\label{central:f,g}
  K(f_1,g_1;f_2,g_2) & =+\frac{r^2}{16\pi}\oint_{\partial\Sigma}d^2x\sqrt{\gamma}\Big[D^Ag_2\big(-\frac{1}{r}D^2g_1+\frac{1}{r^2}D^2f_1+D^2g_1\nonumber\\
  &-\frac{a}{4}\frac{\cos\theta}{\sin^2\theta}\partial_rD^2g_1\big)\delta_{A\theta}+D^2g_2\big(-\frac{r}{2}\partial_rD^2g_1-\frac{a}{r}\frac{\cos\theta}{\sin^2\theta}D^Bg_1\delta_{B\theta}\nonumber\\
  &+\frac{2M}{r^2}f_1-\frac{1}{r}D^2f_1-\frac{1}{2}\partial_rD^2f_1+\frac{5a}{4r^2}\frac{\cos\theta}{\sin^2\theta}\delta_{B\theta}D^Bf_1\big)\nonumber\\
  &-f_2\big(-\frac{1}{r}D^2D^2g_1+\frac{1}{r^2}D^AD^2g_1\delta_{A\theta}+\frac{2M}{r^2}D^2g_1\nonumber\\
  &-\frac{a}{2r^2}\partial_{\theta}(\frac{\cos\theta}{\sin^2\theta})D^2g_1-\frac{a}{4r^2}\frac{\cos\theta}{\sin^2\theta}D^AD^2g_1\delta_{A\theta}-\frac{1}{2}D^2\partial_rD^2g_1\big)\Big]
\end{align}

We can now consider the properties of this algebra. In the first step, we consider the gauge aspects as follows
\begin{equation}\label{g asp: simp}
  f_1=r,~~g_1=0,~~f_2=f, ~~g_2=g
\end{equation}

Using the above assumption~(\ref{g asp: simp}) and replacing them in~(\ref{central:f,g}), we have the central term as follows
\begin{equation}
  K(f_1,g_1;f_2,g_2)=0
\end{equation}
So we conclude that
\begin{equation}\label{H_rS}
   \{H_{r_S,0},H_{f,g}\}=0
\end{equation}

With the above result, Averin~\cite{o} considers the $H_{r_S,0}$ to be corresponding to the ADM-energy subtracted of by the energy passing through future null infinity and the portion of the event horizon between the location of $\partial \Sigma$ and the horizon's future end point. According to~(\ref{H_rS}), he considers the $H_{r_S,0}$ as the same as the ADM-energy and then concludes that the surface degrees of freedom are gapless excitations that they keep the ADM-energy invariant. We can achieve the same results in the Kerr spacetime, and we can consider the surface degrees of freedom that provide soft black hole hair.

We have found the gauge aspects of a Kerr black hole as functions on $S^2$. These surface degrees of freedom (\ref{bms02}) justify the surface charge algebra (\ref{algebra:2}) with respect to the Poisson-bracket. This is an evidence for a theory living on a lower dimension that describes part of the phase space near the Kerr solution $g_{ab}$. This is another realization of the holographic principle \cite{Hft,Skd} for a Kerr black hole spacetime.

Finding the surface degrees of freedom from (\ref{dH:final}) does not mean the microstates as black hole soft hair have zero energy. The energy of the physical microstates of the Kerr black hole can be given by the black hole mass and this is not zero in the Kerr case \cite{av2}. Kerr black holes with non-zero mass $M$ and spin $J$ have a hidden conformal symmetry \cite{Cast}. This symmetry on soft modes that appear on the low-energy theory, can not be found in a near horizon region of spacetime. The symmetry can be found in the near horizon region of the phase space \cite{d2,hac1,hac2,hac}.

We have been able to find the surface degrees of freedom for the Kerr black hole, which are the gauge aspects which
can be defined as some functions on $S^2$.  Utilizing the Poisson bracket, the algebra of the gauge aspects can be given
by (\ref{algebra:2}). Given the surface degrees of freedom and their algebra, we can claim to have arrived at a lower-dimensional
theory that describes part of the phase space. This result is in consistent with the result in \cite{Cast} that suggest existence of a
hidden conformal symmetry that emerges in the near horizon of the phase space. The coordinates of this part of the phase
space are determined by gauge aspects defined on a $2D$ sphere that this is a reason for the lower dimensions of the theory
that describes the near event horizon region.

\section{Conclusion}
\label{sec:5}
We have extended a phase space formalism to find a hidden symmetry in the near horizon region of the Kerr black
hole and to describe a possible Kerr/CFT holography that can be a good tool for description of mechanics of the black hole.
We have presented an approach to find the physical gauge degrees of freedom that can be the responsible for the black
hole microstates. These states are the Bogoliubov gapless modes that their degeneracy can be lifted by the quantum effects.
In this conclusion section, we analyze the results in more details.

The 2D theory on the event horizon that we have found in the last section describes part of the phase space. The part of the phase space is near the reference point $g_{ab}$ of the Kerr spacetime. The gauge aspects are gapless excitations of a Kerr black hole. We guess the 2D theory on the event horizon is a conformal field theory which is related to scale-invariance of the system which include the gapless modes. This symmetry  can be lifted by a conformal anomaly which is seen by the appearance of central extensions in the Virasoro algebra in the calculation of quantum effects. The suggested theory to  describe the near horizon region is a 2D conformal field theory (CFT) where the reason for being
 two-dimensional goes back to the gauge excitations dependences on 2D sphere coordinates, and the reason for being
conformally invariant is due to the scale invariance as mentioned in section \ref{sec:intro}.

A necessary condition for the dual theory to be conformally invariant is that it has a 2D-stress tensor with the Virasoro-algebra. There, the Virasoro-generators are the generators of the gauge transformations and the generators from the gauge aspects satisfy a Virasoro algebra as follows
\begin{equation}\label{Vir01}
\{H_n,H_m\}=(m-n)H_{m+n}+K_{m,n},
\end{equation}
where the bracket $\{,\}$ is the Dirac bracket on the gauge fixed space and the central term $K_{m,n}$ could
have the following relation
\cite{d2,hac1,hac2,hac}
\begin{equation}\label{cent01}
K_{m,n} =\delta H(\xi_n,\mathcal{L}_{\xi_m}g;g),
\end{equation}
where the Hamiltonian generators utilizing the gauge aspects $(f_n,g_n)$ could be as follows
\begin{equation}
  H_n:=H_{(f_n,g_n)}.
\end{equation}

The authors in \cite{hac} write the diffeomorphism vector fields $\xi_n$ using the Kerr metric in the conformal coordinates
and are able to show that these vectors could satisfy two copies of the Witt-algebra as follows
\begin{eqnarray}
  \big[\xi_m,\xi_n\big] &=& -i(m-n)\xi_{m+n}\nonumber \\
  \big[\bar{\xi}_n,\bar{\xi}_m\big] &=& -i(m-n)\bar{\xi}_{m+n}\nonumber \\
  \big[\xi_n,\bar{\xi_m}\big] &=& 0
\end{eqnarray}

Since our initial metric and the metric used by the authors in \cite{d2,hac1,hac2,hac} are the Boyer-Lindquist metric, we can claim that
such vectors can be found in the Bondi coordinate as well to satisfy the With algebra. In the Bondi coordinates and utilizing
the fall-off conditions, we can use the results of the central terms in \cite{d2,hac1,hac2,hac} as follows
\begin{equation}\label{cent02}
K_{m,n}=\frac{1}{12}aMm^3\delta_{m+n}
\end{equation}
where its dependence on the angular momentum can be seen from the $a$-dependence terms in (\ref{dH: in proj-2}). Of course, in order to the central term to be independent of the black hole temperature, the Wald-Zoupas counterterm
must also be added to the Hamiltonian variation, which has been done in \cite{hac}.

Finding the Virasoro-algebra is a good reason why the dual theory we found in the section~\ref{sec:4} could be a conformally  invariant one. This means that the Virasoro-generators made of the surface degrees of freedom could justify the algebra. In other words, the surface degrees of freedom are good candidates for the soft hair of the black hole. Of course, it should be noted that the part of the phase space does not cover the whole Kerr-family. The dual theory that we have found out describes part of the phase space near the reference point and its gapless excitations. The microstates that are candidates the black hole soft hair must not have zero energy. To satisfy this condition, it is needed the Virasoro zero modes $H_0$ and $\bar{H}_0$ that measure the mass and angular momentum of the Kerr black hole \cite{av2} is non-zero. This could be a good sign that the Kerr/CFT duality can be a good candidate to describe the black hole soft hair.

In this way, we have a lower dimensional theory for the Kerr black hole that governs the part of the phase space related to the physical microstates. Is this theory a correct one? What kind of the conformal field theory is considered to describe the Kerr black hole spacetime? Another check that one can do to see whether the theory is a correct candidate is related to the black hole entropy. Utilizing the central
charges, the entropy can be obtained with the help of the Cardy formula. The result obtained in \cite{d2,hac1,hac2,hac} can also be used in the approach we used.

That, the part of the phase space can be responsible for the physical microstates, is obtained from a field theoretic point of view. In a bosonic and self-interactive field theory, when the stationary critical point of the theory has been accompanied by some gapless modes, the degeneracy of the modes can be lifted by the quantum effects \cite{i1}. As the field theoretic states, the Kerr family of the stationary and asymptotically flat solutions of the Einstein gravity are the critical field configurations and these states should have the gapless modes. We have these modes in our analysis that are the physical origins of the black hole entropy.

As the approach utilized in \cite{o} and \cite{av2}, we could deduce the Hamiltonian generators of the diffeomorphisms contains the Virasoro generators of a possible CFT that governs the black hole microstates. We can not find more details on the CFT by this approach. As stated in \cite{av2}, the approach to deduce the existence of the Virasoro generators is very sensitive to the choice of the phase space related to the microstates. In addition, according to the Virasoro algebra (\ref{algebra:2}), it is possible to count the degeneracy of the states. By finding the degeneracy, we can compare it to the expected entropy. If we can go this way, we can deduce the theory covers all degrees of freedom of the Kerr black hole. So far we have been able to present a good candidate to describe the Kerr black hole soft hair and the Kerr/CFT duality. We leave the consistency check on the entropy counting to clarify more the 2D CFT for the future investigations.

\section*{Acknowledgements}
We thank  Artem Averin for insightful discussions and commenting on a draft of this paper.

\appendix
\section{Appendix}\label{app}

In this section, we present the Christoffel coefficients of the Kerr spacetime. We use the following well-known relation
\begin{equation}\label{Christ Gen}
  \Gamma^a_{bc}=\frac{1}{2}g^{ad}(\partial_cg_da+\partial_bg_dc-\partial_dg_bc)
\end{equation}
For $r\to \infty$ we have the following coefficients
\begin{eqnarray}
  \Gamma^v_{vr} &=& \Gamma^V_{v\phi}=\Gamma^v_{ra}=\Gamma^v_{r\theta}=\Gamma^v_{\theta\phi}=0\nonumber \\
  \Gamma^r_{v\phi} &=& \Gamma^r_{r\phi}=\Gamma^r_{\theta\phi}=0\nonumber \\
  \Gamma^{\theta}_{v\phi} &=& \Gamma^{\theta}_{rr}=\Gamma^{\theta}_{r\phi}=\Gamma^{\theta}_{\theta\phi}=0\nonumber \\
  \Gamma^{\phi}_{va} &=& \Gamma^{\phi}_{rr}=\Gamma^{\phi}_{r\theta}=\Gamma^{\phi}_{\theta\phi}=\Gamma^{\phi}_{\phi\phi}=0\nonumber \\
  \Gamma^{\theta}_{vv} &=& \Gamma^{\theta}_{vr}=\Gamma^{\theta}_{\theta v}=O(r^{-4})\nonumber \\
  \end{eqnarray}
which are the coefficients that are either exactly zero or are approximately zero. The non-zero coefficients are as follows
\begin{eqnarray}
  \Gamma^r_{rr} &=& O(r^{-3})\nonumber\\
  \Gamma^r_{AB} &=& rV\gamma_{AB}\nonumber\\
  \Gamma^v_{AB} &=& -r\gamma_{AB}\nonumber\\
  \Gamma^{\phi}_{r\phi} &=& \Gamma^{\theta}_{r\theta}=\frac{1}{r} \nonumber\\
  \Gamma^v_{vv} &=& \Gamma^r_{vv}=\frac{M}{r^2} \nonumber\\
  \Gamma^v_{\theta\theta} &=& -r \nonumber\\
  \Gamma^{\theta}_{\theta\theta} &=& \frac{\cos\theta}{\sin^2\theta}\frac{a}{r}\nonumber\\
  \Gamma^r_{r\theta} &=& -\frac{1}{2}\frac{\cos\theta}{\sin^2\theta}\frac{a}{r}
\end{eqnarray}

In the calculation of the coefficients, we also noted that the general form of the metric also applies to the Bondi gauge.

\end{document}